\documentclass[conference]{IEEEtran}
\IEEEoverridecommandlockouts
\usepackage{cite}
\usepackage{amsmath,amssymb,amsfonts}
\usepackage{algorithmic}
\usepackage{graphicx}
\usepackage{textcomp}
\usepackage{xcolor}
\usepackage{balance}
\usepackage[all]{nowidow}
\usepackage{booktabs}
\usepackage{subfig}

\usepackage[hidelinks]{hyperref}
\usepackage[capitalise]{cleveref}

\begin{document}

\title{Achieving Realistic Cyclist Behavior in SUMO using the SimRa Dataset}

\author{
\IEEEauthorblockN{Ahmet-Serdar Karakaya\IEEEauthorrefmark{1}, Ioan-Alexandru Stef\IEEEauthorrefmark{1}, Konstantin Köhler\IEEEauthorrefmark{1},\\ Julian Heinovski\IEEEauthorrefmark{2}, Falko Dressler\IEEEauthorrefmark{2}, David Bermbach\IEEEauthorrefmark{1}}

\IEEEauthorblockA{\IEEEauthorrefmark{1}\textit{Technische Universit\"at Berlin \& Einstein Center Digital Future}\\
\textit{Mobile Cloud Computing Research Group} \\
\{ask,iast,koko,db\}@mcc.tu-berlin.de}
\IEEEauthorblockA{\IEEEauthorrefmark{2}\textit{Technische Universit\"at Berlin}\\
\textit{Telecommunication Networks}
\{heinovski,dressler\}@ccs-labs.org}
}

\maketitle

\begin{abstract}
Increasing the modal share of bicycle traffic to reduce carbon emissions, reduce urban car traffic, and to improve the health of citizens, requires a shift away from car-centric city planning.
For this, traffic planners often rely on simulation tools such as SUMO which allow them to study the effects of construction changes before implementing them.
Similarly, studies of vulnerable road users, here cyclists, also use such models to assess the performance of communication-based road traffic safety systems.
The cyclist model in SUMO, however, is very imprecise as SUMO cyclists behave either like slow cars or fast pedestrians, thus, casting doubt on simulation results for bicycle traffic.
In this paper, we analyze acceleration, deceleration, velocity, and intersection left-turn behavior of cyclists in a large dataset of real world cycle tracks.
We use the results to improve the existing cyclist model in SUMO and add three more detailed cyclist models and implement them in SUMO.
\end{abstract}

\begin{IEEEkeywords}
Urban planning, Motion sensor data, Sensor data analysis, Traffic simulation
\end{IEEEkeywords}

\section{Introduction}
\label{sec:intro}
Active transportation modes such as cycling provide health benefits, alleviate traffic congestion, and reduce air pollution~\cite{goetschi2016cycling}.
In practice, however, cyclists often face a car-centric traffic infrastructure which has a significant impact on their (perceived) safety and also affects the attractiveness of cycling routes~\cite{karakaya2020simra, pedroso2016bicycle,aldred2018predictors}.
Changing this infrastructure to better accommodate cyclists and pedestrians requires significant planning efforts of city planners and traffic engineers.
Similarly, road traffic safety systems for vulnerable road users are often assessed using simulation.
Particularly the interaction with cars is relevant when it comes to V2X-based safety systems for cyclists~\cite{oczko2020integrating}.
Many of these studies rely on the open source simulation platform SUMO\footnote{https://www.eclipse.org/sumo/} (Simulation of Urban Mobility), which allows them to study the effects of infrastructure changes before implementing them on the streets.

In SUMO, vehicles and their dynamics are simulated individually~\cite{lopez2018microscopic}.
Unfortunately, the cyclist model is not particularly realistic -- cyclists can either be modeled to behave as slow cars or as fast pedestrians.
Several studies have already improved the bicycle model of SUMO.
For instance, Kaths et al.~\cite{kaths2016integration} investigated the intersection behavior of cyclists using camera traces and transferred findings into SUMO.
Also, Grigoropoulos et al.~\cite{grigoropoulos2019modelling} improved modeling of bicycle infrastructure at intersections while Heinovski et al.~\cite{heinovski2019modeling} created a virtual cycling environment to import real bicycle behavior directly into SUMO.
Nevertheless, the current cyclist behavior in SUMO is still rather unrealistic; so far, researchers have devoted much more effort to car models, e.g., \cite{chandler1958traffic,gazis1961nonlinear,gipps1981behavioural, leutzbach1986development,bando1995dynamical,krauss1998microscopic,treiber2000congested,salles2020extending}.
One reason for this is that, until recently, not enough data on real-world cyclist behavior have been available.
Today, crowdsourced data collection approaches such as SimRa\footnote{https://github.com/simra-project/}~\cite{karakaya2020simra} have made thousands of cycle tracks available as open data.

In this paper, we analyze the SimRa dataset regarding acceleration, deceleration, and velocity of cyclists as well as their left-turn behavior in four-way intersections.
We then use our findings to improve the cyclist model in SUMO.
Additionally, we add three more detailed cylist models for slow, medium and fast cyclists.
In this regard, we make the following contributions:
\begin{itemize}
    \item We show that SUMO's default bicycle simulation is not realistic,
    \item we improve bicycle simulation of SUMO by deriving new parameters for that vehicle type in SUMO,
    \item we add three new bicycle simulation models - slow, mediumd and fast - to SUMO by splitting the SimRa dataset into slow, medium and fast rides.
    \item we develop an intersection model which captures cyclists' left-turn behavior at intersections in a more realistic way, and
    \item we compare our improvements to SUMO's default bicycle simulation, using the SimRa dataset as a ground truth.
\end{itemize}

This paper is an extension of our previous work~\cite{karakaya2022realistic}.
The main new contributions are:
\begin{itemize}
\item We rerun all analyses of the original paper using significantly more rides which have become available in the time period since starting our original paper.
\item We complement the general cyclist model from our original paper with three dedicated models covering the behavior of cyclist groups clustered by their average velocity.
    For each of the three models, we also derive distributions of left-turn, acceleration, and maximum velocity behavior.
\item For both the general and the three dedicated cyclist models, we also derive deceleration behavior as a distribution.
\item We implement all new contributions as SUMO plugins and have successfully reimplemented the cyclist model from our original paper in SUMO core.
    We are currently in the process of finalizing an implementation of the extensions of this paper also in SUMO core.
\item We evaluate all four cyclist models against the default cyclist model of SUMO using scenarios from Berlin, Munich, and Hanover, Germany
\end{itemize}

\section{Background}
\label{sec:background}
In this section, we give an overview of SUMO (see~\cref{sub:sumo}) and SimRa (see~\cref{sub:simra}), which provided the dataset we used in our work.

\subsection{SUMO}%
\label{sub:sumo}

SUMO is an open source traffic simulation tool that offers macroscopic as well as microscopic simulation of vehicle mobility~\cite{lopez2018microscopic}.
SUMO includes models for different types of ``vehicles'', including, among others, cars, bicycles, and even pedestrians.
Due to its large feature set, it has become the de-facto standard for traffic simulation and is used even beyond the transport community, e.g., \cite{beilharz2021towards}.

Traffic scenarios are, among other things, defined by road networks and vehicle traffic.
The road network includes roads and their (sub-)lanes as well as exclusive lanes for cyclists and pedestrians, or road-side infrastructure such as traffic lights.
Furthermore, connections between these lanes and traffic lights can be configured.

When modeling vehicle traffic, users specify demand for a specific road segment per vehicle type and can adjust vehicle-specific parameters of SUMO's simulation model to control their respective behavior.
In general, vehicle parameters are usually specified in the vehicle type declaration (\textit{vType}), applying the changes to all instances of the respective \textit{vType}, e.g., to all cars.
An alternative, however, is to obtain multiple \textit{vType} realizations which typically differ in at least one parameter by using so-called \textit{vTypeDistributions}.
This way, when spawning a new vehicle, SUMO randomly picks a specific \textit{vType} from the \textit{vTypeDistribution} and instantiates the vehicle's parameters accordingly, e.g., cars can thus have individual maximum velocities.

In SUMO, vehicle behavior is, among other things, defined by
\textit{Car Following (CF) models} for the longitudinal kinematic behavior,
\textit{Lane Change (LC) models} for the lateral kinematic behaviour,
and \textit{junction models} for the behavior at junctions and intersections.

Despite including several of these models for cars and trucks, SUMO does not provide a dedicated movement model for cyclists.
Instead, cyclists are simulated by modeling them either as slow cars or fast pedestrians.
Both of these approaches use movement models of the corresponding vehicle type and adapt their respective shape and kinematic characteristics (e.g., velocity and acceleration profiles) to match cyclists.
While this is obviously a rough approximation, it is unlikely to reflect the behavior of real-world cyclists~\cite{grigoropoulos2019modelling}.

\subsection{SimRa}%
\label{sub:simra}

SimRa is an open source project started in 2019 which aims to identify hotspots of near miss incidents in bicycle traffic~\cite{karakaya2020simra, temmen2022crowdsourcing}.
For this, the project follows a crowdsourcing approach in which cyclists record their daily rides using a smartphone application available for both Android and iOS.
Today, the project has managed to record almost 90,000 rides, most of them in Germany, approximately half of them in Berlin.

During the ride, SimRa records the GPS trace at 1/3 Hz and the motion sensors, i.e., (linear) accelerometer, gyroscope, and rotation vector at 50 Hz;
the motion sensor readings are aggregated by calculating a moving average with a window size of 30 and then keeping only every fifth value.
This was done for saving memory, battery, and mobile data usage while still being able to reconstruct the ride and detect near miss incidents.
After the ride, SimRa shows the recorded ride as a route on the map which is then annotated, cropped (for privacy reasons), and uploaded by the user.
In this paper, we only use measured data from the ride files and disregard user-annotated data on near miss incidents.

\section{Related Work}
\label{sec:rw}
In this section, we give an overview of related work on improving intersection behavior (\cref{sec:rw_intersection}) and longitudinal (\cref{sec:rw_longitudinal}) behavior of cyclists in SUMO's simulation models.

\subsection{Intersection Behavior of Cyclists}%
\label{sec:rw_intersection}

Kaths et al.~\cite{kaths2016integration} aim to address the shortcomings of SUMO's intersection model for cyclists.
For this, they record video footage of an example intersection in Munich and derive cyclist trajectories.
From the set of trajectories, they select one representative trajectory for each combination of start and end points in the intersection and make it available to SUMO via an external API.
While this is a significant improvement in realism over SUMO's intersection model, it is hard to generalize to other intersections and cannot cover the plurality of trajectories chosen by real-world cyclists.

Similar to Kaths et al.~\cite{kaths2016integration}, Grigoropoulos et al.~\cite{grigoropoulos2022traffic} analyze video footage of intersections with the goal of better understanding the intersection behavior of cyclists.
Their focus, however, is not on deriving an improved intersection model but rather on identifying best practices for traffic planners working on intersections with high volumes of cycling traffic.
Grigoropoulos et al.~\cite{grigoropoulos2019modelling} propose to adjust the default traffic infrastructure inside SUMO to achieve more realistic cyclist behavior at intersections.
Here, they focus on the number and shape of bicycle lanes which, however, are highly specific and differ from intersection to intersection.

\subsection{Longitudinal Behavior of Cyclists}%
\label{sec:rw_longitudinal}

Twaddle et al.~\cite{twaddle2016modeling} examine four models for the longitudinal kinematic behavior of cyclists, i.e., acceleration and velocity.
The first, called Constant Model, is the simplest and is the SUMO default:
Cyclists accelerate and decelerate at a constant rate until the desired velocity is reached.
This model works well when breaking to a full stop but leads to frequent acceleration jumps between a fixed positive or negative value and zero, which is not realistic cyclist behavior.
In the Linear Decreasing Model, maximum acceleration is reached when starting the acceleration maneuver and then linearly declines until the desired velocity is reached.
This model is outperformed by all other models.
In the third and fourth models, Polynomial and Two Term Sinusoidal Model, acceleration or deceleration start at zero and then gradually grow over time.
In their paper, Twaddle et al.~\cite{twaddle2016modeling} analyze the video recordings of 1,030 rides in four intersections in Munich, Germany and conclude that the Polynomial Model has overall the most realistic cyclist behavior but is, however, not trivial to implement in SUMO.

A different approach of achieving realistic cycling behavior in SUMO is taken by Heinovski et al.~\cite{heinovski2019modeling}.
The authors simulate multiple traffic scenarios in which accidents between cars and cyclists occur to investigate the effects of wireless communication between cyclists and other road users in the context of accident prevention.
In order to obtain realistic cycling behavior for SUMO, they set up a novel Virtual Cycling Environment (VCE) featuring an actual bicycle that is connected to the simulation via multiple sensors.
The VCE supports interactive empirical studies in a physically safe environment and allows the authors to record the cyclists' behavior in the form of trajectories.
They use a set of recorded trajectories from different cyclists for emulating realistic cycling behavior inside SUMO to simulate accidents.
Although their approach produces trajectories from cyclists created with an actual bicycle, it does only achieve limited realism, since no other road users were present when recording the trajectories.
Furthermore, deriving a realistic set of trajectories requires a large number of test persons.

\section{Cycling Behavior in SimRa and SUMO}
\label{sec:analysis}
In this section, we analyze real-world cyclists' behavior extracted from the SimRa dataset and compare it to the behavior of SUMO's default bicycle model.
SimRa's dataset stems from crowdsourced smartphone data generation and thus suffers from poor sensor quality~\cite{chowdhury2014estimating, usami2018bicycle} as well as heterogeneous hardware and users~\cite{basiri2018impact}.
This leads to a lot of unclean data, which we first need to filter out (\cref{sec:analysis_preprocessing}).
Since we want to create one general type for all cyclists and complement it with three dedicated models for slow, medium, and fast cyclists, we need to derive three distinct types from the SimRa dataset, which we do in \cref{sec:analysis_types}.
We then analyze acceleration, deceleration and velocity behavior for each of the four models in \cref{sec:analysis_acceleration,sec:analysis_deceleration,sec:analysis_velocity} before discussing left-turn behavior at four-way intersections of the different models in \cref{sec:analysis_pathfinding}.
We omit a detailed discussion of the right-turn behavior at intersections, since SUMO's default model does not deviate much from the behavior observed from SimRa's dataset.
When referring to SUMO's bicycle model, we refer to the ``slow car'' model of SUMO as the ``fast pedestrian'' model occasionally leads to poor results and was therefore not considered further.

\begin{table}
\centering
\caption{Most important attributes of entities of the SimRa dataset}%
\label{tab:dataset}
\begin{tabular}{ccc}
\toprule
& Total & Used \\
\midrule
\midrule
Number of Rides & 60,470 & 55,175 \\
Number of Accelerations & 12,382,675 & 2,330,292 \\
Number of Decelerations & 12,985,955 & 2,623,904 \\
\bottomrule&
\end{tabular}
\end{table}

Aside from the public SimRa datasets~\cite{dataset_simra_set1,dataset_simra_set2, dataset_simra_set3} and more recent rides available on GitHub,\footnote{https://github.com/simra-project/dataset} we also used non-public rides which have, for privacy reasons, not been published yet.
\Cref{tab:dataset} summarizes the most important attributes of the dataset that we used.

We used rides from almost 100 SimRa regions when calculating the distributions for the maximum acceleration, maximum deceleration and maximum velocity of cyclists.
\subsection{Preprocessing}
\label{sec:analysis_preprocessing}
To achieve the best possible data quality, we tested various pre-processing techniques and filters.
We also conducted an experiment in which sample trajectories were recorded in parallel on several SimRa client devices and compared to a ground truth trajectory recorded by a stand-alone GPS receiver.
In the end, we used a Gaussian Kernel filter for improving location data and a Low Pass filter for the velocity data.
Additionally, the SimRa dataset contains information about the location accuracy, which we use to filter out rides where the GPS accuracy suffered greatly.
After filtering semantically and syntactically defective files, we used data from 55,175 rides, which is around 65\% of the initial dataset as input for our analysis scripts.\footnote{https://github.com/simra-project/SimRaXSUMO}

\subsection{Categorizing Cyclists by Velocity}%
\label{sec:analysis_types}

To get different types of cyclists, we decided to analyze the average velocity of each ride after filtering out the stops.
\Cref{fig:analysis_avg_vel_all} shows the distribution of the average velocities of each ride in the SimRa dataset.
This does not present an obvious way to split the dataset into three types, which is why we have split the dataset from a SUMO users' perspective.
We, hence, decided to split the dataset so, that the {25\% slowest and the 25\% fastest rides represent the slow and the fast cyclists respectively, which leaves the middle 50\% to the medium-paced cyclists.
This results in the following cyclist type velocities:
\textit{Slow cyclists} have an average cycling velocity of up to 13.5 km/h.
\textit{Medium cyclists} have an average cycling velocity between 13.5 km/h and 17.9 km/h.
\textit{Fast cyclists} have an average cycling velocity above 17.9 km/h.
We have chosen against an equal split and in favor of a 25\%-50\%-25\% split, because we think that the majority of the cyclists should be in the same cyclist type, namely, the medium cyclist type.
\begin{figure}
  \centering
    \includegraphics[width=\columnwidth]{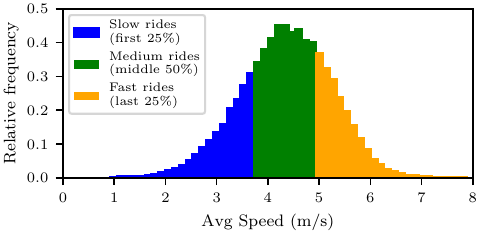}
    \caption{%
        Histogram of the empirical average velocity capabilities of cyclists found in the SimRa dataset. The average velocity of all bicycle rides after the preprocessing is 4.38 m/s with a standard deviation of 0.89 and a median of 4.42 m/s.
    }%
    \label{fig:analysis_avg_vel_all}
\end{figure}

\subsection{Acceleration}%
\label{sec:analysis_acceleration}

\begin{figure*}[t]
    \centering
    \subfloat[All cyclists]{\includegraphics[width=\columnwidth]{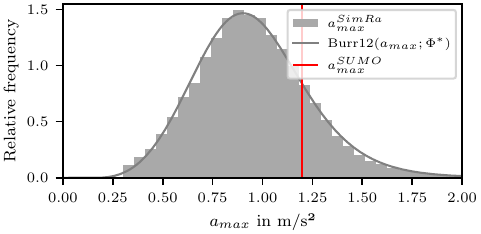}\label{fig:analysis_max_acceleration_dist_fit_all}}
    \hfill
    \subfloat[3 cyclist groups: slow ($s$), medium ($m$), and fast ($f$)]{\includegraphics[width=\columnwidth]{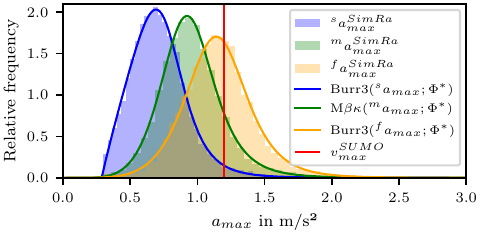}\label{fig:analysis_max_acceleration_dist_fit_groups}}
    \caption{%
        Histogram of maximum acceleration capabilities found in the empirical SimRa dataset and their respective distribution functions.
        The red scalar represents the default value in SUMO.
    }%
    \label{fig:analysis_max_acceleration_dist_fit}
\end{figure*}

For analyzing cyclist acceleration, we extracted acceleration maneuvers from the dataset.
For this, we slightly adapted the approach of \cite{ma2016modeling}.
First, we split the velocity profile at its local extrema to get segments where the cyclist accelerates/decelerates.
Then, we consider only segments with a distance from 20 m to 350 m and a duration between 5 s to 40 s.
We also make sure to filter out segments where the variance in velocity is too low, i.e., such that $ \frac{|v_{s} - v_{e}|}{max(v_{s},v_{e})} > 0.5 $, where $v_{s}$ and $v_{e}$ are the velocities at the start and end of a segment.
With that approach we found 228,347 acceleration maneuvers in the cleaned dataset.
Distribution fitting processes, which were done with SciPy\footnote{https://scipy.org/}, showed that the Burr (Type XII) distribution~\cite{burr1942cumulative} $Burr12(x; c,d) = c*d*\frac{x^{c-1}}{(1+x^c)^{d+1}}$ for $x = a_{max} \geq 0$ and $c,d > 0$  fits the data best for the general cyclist model (see also \cref{fig:analysis_max_acceleration_dist_fit_all}).
For the models of the slow and fast cyclist models, the Burr (Type III) distribution~\cite{burr1942cumulative} 

$Burr3(x; c,d) = c*d*\frac{x^{-c-1}}{(1+x^c)^{d+1}}$ for $x = a_{max} \geq 0$ and $c,d > 0$ fits the data best, while the Mielke Beta-Kappa distribution~\cite{mielke1973distribution} 

$M\beta\kappa(x; c,d) = \frac{k*x^{k-1}}{(1+x^s)^{1+k/s}}$ for $x = a_{max} > 0$ and $k,s > 0$ is the best fit for the medium cyclist model's acceleration distribution (see also \cref{fig:analysis_max_acceleration_dist_fit_groups}).

Comparing the acceleration capability of actual cyclists (the SimRa dataset) with the default SUMO cyclist model, differences become apparent.
By default, SUMO specifies $a_{max}^{SUMO}$ with 1.2 m/s².
This deviates significantly from the findings in the SimRa dataset where only 15\% of the acceleration maneuvers are executed with a maximum acceleration of 1.2 m/s² or higher.
Furthermore, the empirical distributions are rather wide, indicating a broad variance across different cyclist types and cycling situations, which is in stark contrast to SUMO's strategy of choosing a fixed maximum value.

\subsection{Deceleration}%
\label{sec:analysis_deceleration}

\begin{figure*}[t]
    \centering
    \subfloat[All cyclists]{\includegraphics[width=\columnwidth]{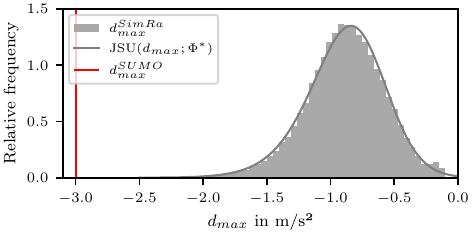}\label{fig:analysis_max_deceleration_dist_fit_all}}
    \hfill
    \subfloat[3 cyclist groups: slow ($s$), medium ($m$), and fast ($f$)]{\includegraphics[width=\columnwidth]{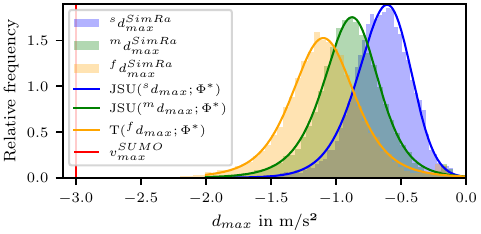}\label{fig:analysis_max_deceleration_dist_fit_groups}}
    \caption{%
        Histogram of maximum deceleration capabilities found in the empirical SimRa dataset and their respective distribution functions.
        The red scalar represents the default value in SUMO.
    }%
    \label{fig:analysis_max_deceleration_dist_fit}
\end{figure*}

For analyzing cyclist deceleration, we extracted deceleration maneuvers from the dataset (see \cref{sec:analysis_acceleration}).
Distribution fitting processes showed here that the Johnson's $S_{U}$-distribution~\cite{johnson1949systems} $JSU(x; a,b) = \frac{b}{\sqrt{x^2+1}}\phi(a+b*log(x+\sqrt{x^2+1}))$ for $x=d_{max}$ and $b>0$ with $\phi$ being the probability density function of the normal distribution, fits the data best for the general cyclist model (see also \cref{fig:analysis_max_deceleration_dist_fit_all}).
$JSU(d_{max}; a,b)$ was also the best fit for the slow and medium cylist models, whereas the fast cyclists' data was the best fit for the Student's $t$-distribution~\cite{student1908probable} $t(x; \nu)=\frac{\Gamma((\nu+1)/2)}{\sqrt{\pi*\nu}*\Gamma(\nu/2)}*(1+x^2/\nu)^{-(\nu+1)/2}$ for $x=d_{max}$ (see also \cref{fig:analysis_max_deceleration_dist_fit_groups}).

Comparing the deceleration capability of actual cyclists (the SimRa dataset) with the default SUMO bicycle model, differences become apparent.
By default, SUMO specifies $d_{max}^{SUMO}$ with -3 m/s².
This deviates extremely from the findings in the SimRa dataset where none of the deceleration maneuvers are executed with a maximum acceleration of -3 m/s² or lower.
Furthermore, the empirical distributions are rather wide, indicating a broad variance across different cyclist types and cycling situations, which is in stark contrast to SUMO's strategy of choosing a fixed maximum value.

\subsection{Velocity}%
\label{sec:analysis_velocity}

\begin{figure*}[t]
    \centering
    \subfloat[All cyclists]{\includegraphics[width=\columnwidth]{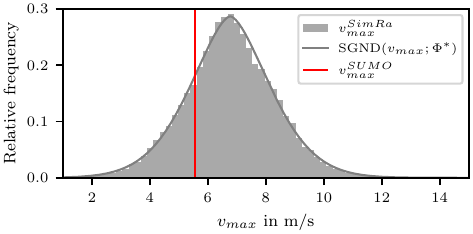}\label{fig:analysis_max_velo_dist_fit_all}}
    \hfill
    \subfloat[3 cyclist groups: slow ($s$), medium ($m$), and fast ($f$)]{\includegraphics[width=\columnwidth]{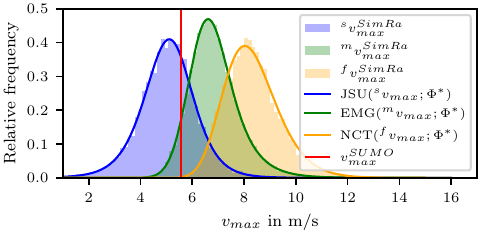}\label{fig:analysis_max_velo_dist_fit_groups}}
    \caption{%
        Histogram of maximum velocity capabilities found in the empirical SimRa dataset and their respective distribution functions.
        The red scalar represents the default value in SUMO.
    }%
    \label{fig:analysis_max_velo_dist_fit}
    \vspace{-.5em}
\end{figure*}

To gain insights into cyclists' behavior regarding their velocities, we calculate the maximum velocity for each ride file in the cleaned SimRa dataset.
Using distribution fitting, we found that the symmetric generalized normal distribution~\cite{nadarajah2005generalized} $SGND(x;\beta) = \frac{\beta}{2*\Gamma(1/\beta)}*exp(-|x|^\beta)$ for $x=v_{max}$ fits the empirical data of all cyclists best (see also \cref{fig:analysis_max_velo_dist_fit_all}) and is therefore a valid fit for the specification of the empirical distribution of $v_{max}^{SimRa}$ for the general cyclist model.
For the model of the slow cyclists, the Johnson's $S_{U}$-distribution~\cite{johnson1949systems} $JSU(v_{max}; \Phi\mbox{*})$ fits the data best, while the best fit for the medium cyclists' model is the Exponentially Modified Gaussian distribution~\cite{grushka1972characterization} $EMG(x; K) = \frac{1}{2*K}*exp(\frac{1}{2*K^2}-x/K)erfc(-\frac{x-1/K}{\sqrt{2}})$ for $x = v_{max}$ and $K>0$.
Finally, the non-central $t$-distribution~\cite{hogben1961moments} $NCT\newline(v_{max}; \Phi\mbox{*})$ emerged as the best fit for the fast cyclists' data.

On the other hand, SUMO sets $v_{max}^{SUMO}$ at 5.56 m/s by default.
This deviates significantly from the findings in the SimRa dataset where 77\% of the rides have a higher maximum velocity.
Bringing this together with the acceleration findings, real-world cyclists often (but not always) cycle much faster than SUMO cyclists and vary much more in their acceleration behavior.

\subsection{Left-turn Behavior at Intersections}%
\label{sec:analysis_pathfinding}

\begin{figure}[t]
    \centering
    \includegraphics[width=\columnwidth]{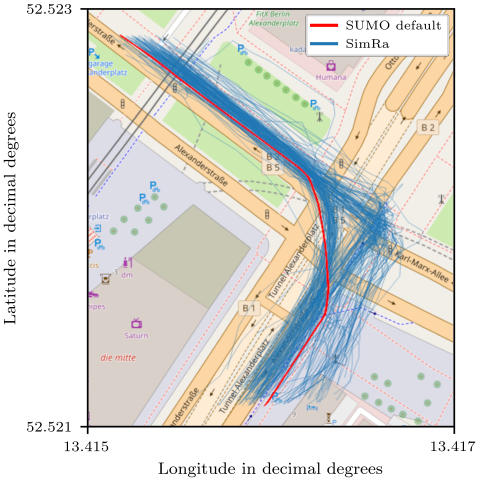}
    \caption{%
        Qualitative comparison between the SUMO default intersection model and real world data given by SimRa for the intersection between Alexanderstraße and Karl-Marx-Allee in Berlin.
        SimRa shows two distinct left-turn paths (i.e., a direct and an indirect one) whereas SUMO default only models the direct path.
    }%
    \label{fig:analysis_im_traj_default}
\end{figure}

According to the SimRa dataset, cyclists either behave like cars (using the normal road) or pedestrians (using the pedestrian crossing) to take left-turns at intersections.
We call the former a \textit{direct} left turn and the latter an \textit{indirect} left turn.

SUMO's default model only provides cyclists with an unrealistic "bicycle-lane-to-bicycle-lane-left-turn" (see also \cref{fig:analysis_im_traj_default}), where the cyclist enters the intersection from a bicycle lane, crossing all car lanes and directly entering the bicycle lane again.

Taking a closer look at real world intersections in the SimRa dataset revealed that there are mainly two intersection types.
In the first intersection type, the \textit{indirect} path is chosen with a probability of 61\% when all cyclists are considered, while medium cyclists and fast cyclists prefer the indirect path with a probability of 87\% and 50\% respectively.
However, on the second intersection type, almost all cyclists choose the \textit{indirect} path.
Slow cyclists almost never tend to do direct left-turns, since they presumably avoid car traffic the most.
Randomly investigating intersections of both types revealed that the first intersection type has no specific characteristics while the second intersection type actively encourages cyclists to indirect turns through the design of the intersection, e.g., by having a traffic island in the center.
Since such information cannot be identified in OpenStreetMap data reliably and in an abstract way, we will consider only the first intersection type in the following.

SUMO and real world data differ decisively in all metrics considered, namely acceleration, velocity, and left-turn behavior at intersections.
However, these three metrics are crucial for realistically simulating bicycle traffic.
In the following, we try to adapt SUMO to simulate a more realistic cyclist behavior by introducing three different cyclist types.

\section{Improving SUMO's Bicycle Simulation}
\label{sec:concept}
To improve the simulation, we propose three changes to SUMO's bicycle model:
First, the longitudinal kinematic parameters of SUMO's default bicycle model are \mbox{(re-)}parameterized based on the findings from the SimRa dataset.
Second, a novel simulation model is derived from SimRa trajectories to exclusively simulate realistic left-turn bicycle behavior at intersections based on the findings in \cref{sec:analysis}.
The latter model is referred to as the intersection model in the following.
Third, four different cyclist models, with different longitudinal kinematic and left-turn behaviors depending on the findings from \cref{sec:analysis}.
One model is for all cyclists combined and the three other models are for modeling slow, medium and fast cyclists.

\subsection{Longitudinal Kinematic Behavior}%
\label{sec:concept_param}

In \cref{sec:analysis}, we derived maximum acceleration, maximum deceleration and maximum velocity characteristics from the SimRa dataset.
We now use them to improve the longitudinal kinematic behavior of the default SUMO bicycle model.
Contrary to the default parameterization, we use theoretical distribution functions instead of scalar values for the exposed kinematic parameters.
This enables the model to produce more realistic bicycle simulation results since the heterogeneity of real world cycling styles is reflected.

We derive the theoretical distributions by aggregating the respective features from \cref{sec:analysis_acceleration,sec:analysis_deceleration,sec:analysis_velocity}.
For this, we rely on the \textit{law of large numbers} which states that the average of the results obtained from a large number of trials of the same experiment eventually converges to its true expected value~\cite{etemadi1981elementary}.
In the context of this work, this means that individual rides do not matter but that the aggregates of multiple rides will converge towards their actual expected value given a sufficiently large number of rides.

For the implementation, we used \textit{vTypeDistributions} following the results of our previous analysis and sample both distribution independently.

It should be noted that through the parameterizations with theoretical probability density functions SUMO's \textit{speedDev} parameter becomes obsolete as variance between the kinematic preferences among cyclists are already represented by the distribution function.

Furthermore, alternatives for the acceleration parameters have been added to SUMO. This would enable a user to pick a normal distribution and choose the parameters for it.
This is would yield more realistic scenario data contrary to using a simple scalar value for the acceleration.

\subsection{Left-turn Behavior at Intersections}%
\label{sec:concept_im}

To improve the degree of realism in cyclists' left-turn behavior at signaled intersections, we use an adapted version of the external intersection model (a Python script that steers cyclists via SUMO's \textit{Traffic Control Interface}) as proposed by Kaths et al.~\cite{kaths2016integration} which is based on previously recorded real-world trajectories as their guidelines for cyclists across a single predefined intersection.
Our approach algorithmically synthesizes the cyclists' trajectories (i.e., their respective guidelines across the intersection) for any regular four-way intersection and can therefore be seen as a step towards a more universal solution.

The left-turn maneuver distribution, as we call it, specifies the probability of the cyclists choosing either the \textit{direct} or the \textit{indirect} path to cross the intersection.
For this, we use the distributions derived in \cref{sec:analysis_pathfinding} as the default for our intersection model for each different cyclist group.
Users, however, can adjust the distributions if desired or needed for their specific purposes (see also the exception cases in \cref{sec:analysis_pathfinding}).

In order to simplify the process and to improve performance, we decided to integrate this feature into the SUMO core.
The implementation provides a new parameter that lets the user adjust the indirect left turn probability of a bicycle ride.
To detail the workflow, the cyclist makes a decision before each intersection, where there is at least one direct and indirect left-turn available to choose from.
This delegates decisions inside SUMO just by using the bicycle vehicle type parameter.

\subsection{Different Cyclist Models}%
\label{sec:concept_types}

Our approach defines distibutions of models derived from the same default implementation of SUMO.
The models vary only in the parameter values used.
The parameters considered are maximum acceleration (\textit{accel}), maximum deceleration (\textit{decel}) and maximum velocity (\textit{maxSpeed}), splitted into 3 groups, as it can be seen in \cref{sec:analysis_acceleration,sec:analysis_deceleration,sec:analysis_velocity}.
In order to leverage distribution parameters for a bicycle model, we used the \textit{vTypeDistribution} from SUMO to define a distribution of cyclists having all the parameters mentioned.

With this data ready, the only remaining step is to augment these groups with a left-turn behavior probability per group.
In order to make a decision whether to do a direct or an indirect left-turn, we approached the problem by using a script that controls the cyclists, as described in \cref{sec:concept_im}.

Thus, this group-based model enables the user to simulate bicycle traffic in SUMO more realistically.

\section{Evaluation}
\label{sec:eval}
In this section, we evaluate the new cyclist models from \cref{sec:concept} by comparing them to each other and to SUMO's default simulation model and the real-world data taken from the SimRa data set.
We start by introducing the simulation setup (\cref{sec:eval_setup}) which we used to create and run the scenarios with.
We then continue analyzing acceleration (\cref{sec:eval_accel}), deceleration (\cref{sec:eval_deceleration}), velocity (\cref{sec:eval_velocity}), and left-turn behavior at intersections (\cref{sec:eval_turn}) before evaluating the combination of all model extensions (\cref{sec:eval_combined}).
Please note: While it may appear obvious that using the SimRa data set for both parameterization and evaluation should lead to perfect results, this is not the case as our extensions are subject to the design restrictions imposed by SUMO

\subsection{Simulation Setup}%
\label{sec:eval_setup}

As SUMO users can import real-world scenarios from OSM data, simulation results can be compared to real-world data and thus be evaluated.
For our evaluation, we chose specific traffic scenarios from multiple SimRa regions that are representative and likely to showcase both strengths and weaknesses of our extensions.
Likewise, we do not show every cyclist type's results in detail to avoid too many figures and tables.
For evaluating the longitudinal behavior (acceleration, deceleration and velocity), we chose urban traffic scenarios with long straight sections.
As example locations, we use \textit{Oranienstraße} in Berlin, \textit{Dachauer Straße} in Munich and \textit{Frauentorgraben} in Nuremberg.
For evaluating left-turn behavior, we chose compact scenarios around signaled intersections with multiple lanes on each axis.
For this, we study three intersections in Berlin, namely at \textit{Mehringdamm} (see also \cref{fig:mehringdamm_sumo}), \textit{Warschauer Straße}, and \textit{Alexanderstraße}.

\begin{table}%
\centering
\caption{Most important attributes of our simulation scenarios.}%
\label{tab:scenarios}
\resizebox{\columnwidth}{!}{
\begin{tabular}{ccccc}%
\toprule%
                & Cars  & Bicycles & Distance & Lanes\\%
\midrule%
\midrule%
Oranienstr.     & 3,158 & 300      & 1,528 m  & --\\%
Dzchauer Str.   & 3,427 & 300      & 1,204 m  & --\\%
Frauentorgraben & 1,625 & 300      & 1,023 m  & --\\%
Mehringdamm     & 180   & 2,700    & --       & 29\\%
Warschauer Str. & 180   & 2,700    & --       & 26\\%
Alexanderstr.   & 180   & 2,700    & --       & 25\\%
\bottomrule&%
\end{tabular}%
}
\end{table}

We made our scenarios as realistic as possible by choosing main roads and intersections with a lot of traffic and also added a significant number of cars, as \cref{tab:scenarios} shows.
\begin{figure}
    \centering
    \includegraphics[width=\columnwidth]{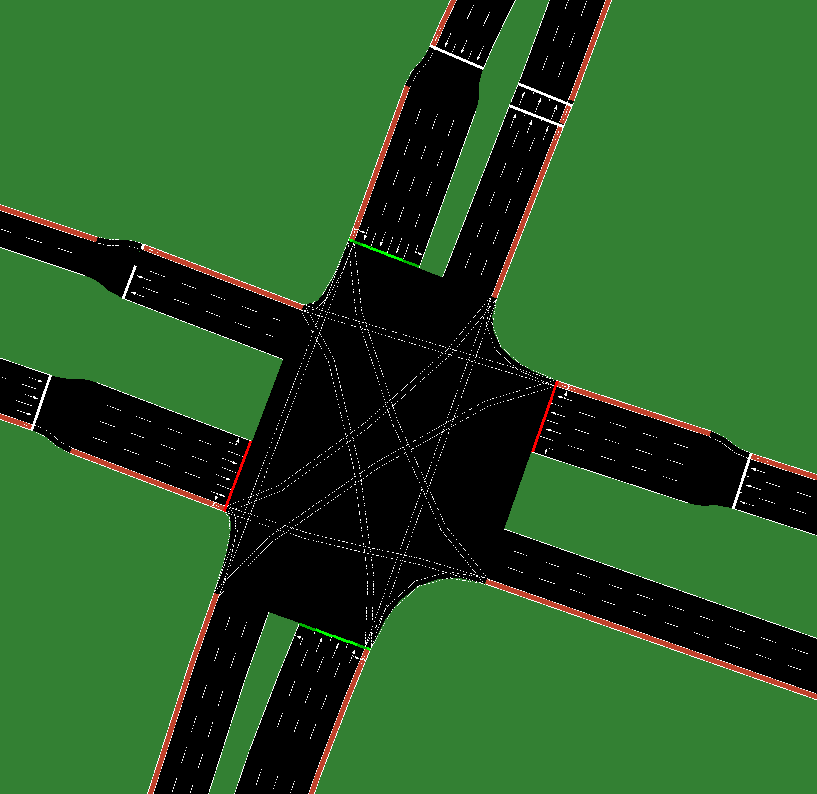}
    \caption{%
        Excerpt from the \textit{Mehringdamm} scenario in SUMO.
        The scenario was created using OSM data only.
    }%
    \label{fig:mehringdamm_sumo}
\end{figure}

For our evaluation, we use SUMO version 1.14.0 and a step size of 1 s in simulations.
The SUMO default results are obtained with SUMO's \textit{vType} \texttt{Bicycle} for cyclists, i.e., the maximum acceleration, maximum deceleration and maximum velocity are scalars and set to 1.2 m/s², -3 m/s² and 5.56 m/s respectively.

\subsection{Acceleration}%
\label{sec:eval_accel}

\begin{figure*}[t]
    \vspace{-1.5em}
    \centering
    \subfloat[All cyclists]{\includegraphics[width=\columnwidth]{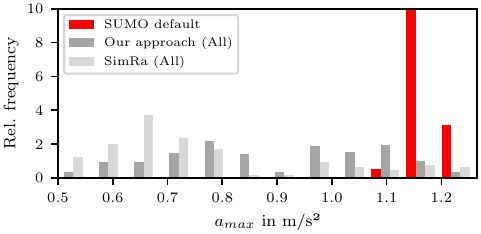}\label{fig:eval_acc_all}}
    \hfill
    \subfloat[Only cyclist group slow ($s$)]{\includegraphics[width=\columnwidth]{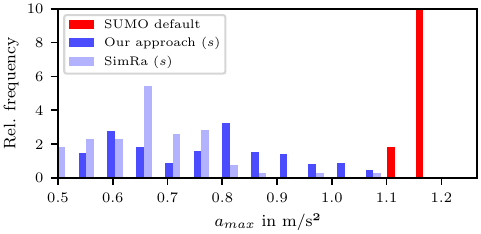}\label{fig:eval_acc_slow}}
    \caption{%
        Histogram of SUMO's, SimRa's, and our approach's observed maximum accelerations inside the example \textit{Frauentorgraben} scenario.
        While the maximum accelerations are heterogeneously distributed in the real-world data and our approach, the default values are clustered.
        Here, the observed accelerations deviate from the configured default value (1.2 m/s²) due to traffic effects inside the simulation.
    }%
    \label{fig:eval_acc}
\end{figure*}

\cref{fig:eval_acc_all} shows the empirical distributions of the maximum acceleration among all cyclist types inside the \textit{Frauntorgraben} scenario simulation and the corresponding real world data.
It is evident that real-world acceleration maneuvers show heterogeneous maximum rates of acceleration.
The same is true for the three cyclist groups of our new approach as \cref{fig:eval_acc_slow} shows.
Note that the other two cyclist groups, namely slow and medium, allow the same conclusion and, that we have randomly chosen the slow cyclist group to visualize in \cref{fig:eval_acc_slow} to make the figure more readable.
Apparently, the default parameterization is not suitable to describe this acceleration behavior among cyclists, as it provides homogeneous maximum acceleration rates within the simulation.
Our new parameterization - for all cyclists together and for each of the cyclist types - is significantly closer to the real-world behavior in the SimRa data set with its highly heterogeneous behavior across cyclists.
\begin{table}
\centering
\caption{Results Overview Maximum Acceleration Comparing SUMO Default, Our Approach (All), Our Approach (Slow), SimRa (All) and SimRa (Slow). All values are in m/s².}%
\label{tab:results_overview_acc}
\begin{tabular}{cccc}
\toprule
& Mean & Std. Deviation & Median\\
\midrule
SUMO default & 1.178 & 0.019 & 1.18 \\
\midrule
Our approach (\textit{all}) & 0.921 & 0.218 & 0.95 \\
SimRa (\textit{all}) & 0.761 & 0.217 & 0.71 \\
\midrule
Our approach (\textit{s}) & 0.727 & 0.168 & 0.74 \\
SimRa (\textit{s}) & 0.656 & 0.109 & 0.65 \\
\bottomrule&
\end{tabular}
\end{table}
This can also be seen, when comparing the mean, standard deviation and median values of SUMO default, all and slow (\textit{s}) cyclists (both in the SimRa dataset and our approach) depicted in \cref{tab:results_overview_acc}.
It also shows, that the division of the cyclist into subgroups further increases the realism, since the difference between the mean and median values of our approach (\textit{all}) and Simra (\textit{all}) shrinks, compared to our approach (\textit{s}) and Simra (\textit{s}).
The standard deviation suffers though, which is probably due to less rides being regarded, because of the splitting into three subgroups.
Please note also here, that we omitted the medium and fast cycling groups to stay consistent with \cref{fig:eval_acc_slow} and avoid clutter.
That our new parameterizations are not a perfect fit indicates that there are probably additional influence factors, e.g., the traffic density or the weather situation, not covered in our kinematic models which aggregate data from all SimRa rides.

\subsection{Deceleration}%
\label{sec:eval_deceleration}

\begin{figure*}[t]
    \centering
    \subfloat[All cyclists]{\includegraphics[width=\columnwidth]{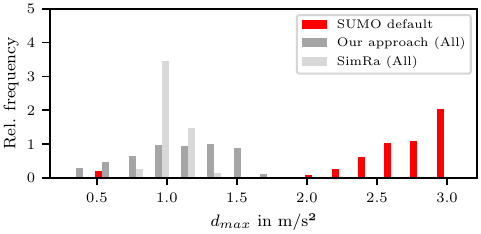}\label{fig:eval_dec_all}}
    \hfill
    \subfloat[Only cyclist group medium ($m$)]{\includegraphics[width=\columnwidth]{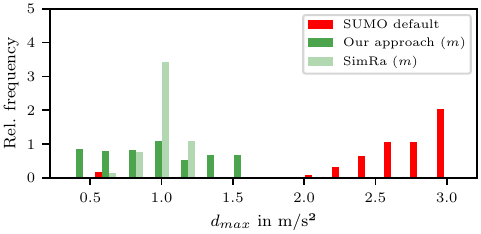}\label{fig:eval_dec_medium}}
    \caption{%
        Histogram of SUMO's, SimRa's, and our approach's observed maximum decelerations inside the example \textit{Dachauer Straße} scenario.
        While the maximum decelerations are heterogeneously distributed in the real-world data and our approach, the default values are clustered.
        Here, the observed decelerations deviate from the configured default value (3m/s\textsuperscript{2}) due to traffic effects inside the simulation.
    }%
    \label{fig:eval_dec}
\end{figure*}

\cref{fig:eval_dec_all} shows the empirical distributions of the maximum deceleration among all cyclist types inside the \textit{Dachauer} \textit{Straße} scenario simulation and the corresponding real world data.
Here, we only show the medium cyclist group as an example to avoid clutter, the other two cycling groups (slow and fast) show very similar results.
It is evident that real-world deceleration maneuvers show heterogeneous maximum rates of deceleration.
The same is true for the medium cyclists as \cref{fig:eval_dec_medium} shows.
Here, too, the default parameterization is not suitable to describe this deceleration behavior among cyclists, as it provides homogeneous maximum deceleration rates within the simulation.
Our new parameterization - for all cyclists together and for each of the cyclist types - is significantly closer to the real-world behavior in the SimRa data set with its highly heterogeneous behavior across cyclists.
\begin{table}
\centering
\caption{Results Overview Maximum Deceleration Comparing SUMO Default, Our Approach (All), Our Approach (Medium), SimRa (All) and SimRa (Medium). All values are in m/s².}%
\label{tab:results_overview_dec}
\begin{tabular}{cccc}
\toprule
& Mean & Std. Deviation & Median\\
\midrule
\midrule
SUMO default & 2.664 & 0.507 & 2.77 \\
\midrule
Our approach (\textit{all}) & 1.071 & 0.336 & 1.09 \\
SimRa (\textit{all}) & 0.962 & 0.141 & 0.96 \\
\midrule
Our approach (\textit{m}) & 0.949 & 0.362 & 0.93 \\
SimRa (\textit{m}) & 0.973 & 0.100 & 0.98 \\
\bottomrule&
\end{tabular}
\end{table}
Just like with the maximum acceleration, \cref{tab:results_overview_dec} shows, that our approach is not only more realistic than SUMO's default bicycle model (both with all and medium (\textit{m}) cyclists), but the introduction of the cyclist groups increases the realism.
The medium cyclist group was randomly chosen as a representative for the other groups, since the conclusion does not differ.
Influence factors, such as the traffic density or the weather situation, not covered in our kinematic models which aggregate data from all SimRa rides, result in a non-perfect fit for our new parameterizations.
Note, that we had to filter out deceleration values, that we deemed too high (above 7 m/s²), to avoid recording emergency deceleration, which is another parameter in SUMO's vehicle parameterization.

\subsection{Velocity}%
\label{sec:eval_velocity}

\begin{figure*}[t]
    \centering
    \subfloat[All cyclists]{\includegraphics[width=\columnwidth]{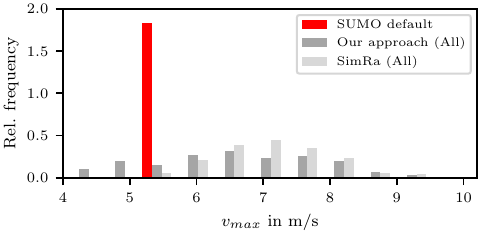}\label{fig:eval_velo_all}}
    \hfill
    \subfloat[Only cyclist group fast ($f$)]{\includegraphics[width=\columnwidth]{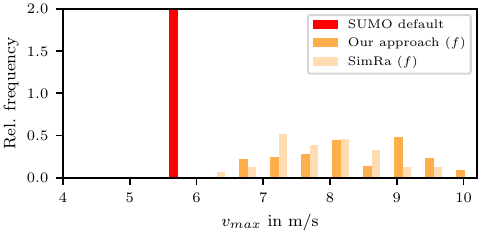}\label{fig:eval_velo_fast}}
    \caption{%
        Histogram of SUMO's, SimRa's, and our approach's observed maximum velocities inside the example \textit{Oranienstraße} scenario.
        While the maximum velocities are heterogeneously distributed in the real-world data and our approach, the default values are clustered.
    }%
    \label{fig:eval_velo}
\end{figure*}

\cref{fig:eval_velo_all} shows the empirical distributions of all cyclists' maximum velocities in the \textit{Oranienstraße} scenario simulation and the real-world scenario.
As with maximum acceleration and deceleration rates, maximum velocities vary widely among real-world cyclists and we only depict one cyclist group (this time fast cyclists) as an example for other groups to increase the readability of the figure.
Once more, the default parameterization is not able to reflect this characteristic.
This is also the case for the different cyclist groups as \cref{fig:eval_velo_fast} shows exemplary.
\begin{table}
\centering
\caption{Results Overview Maximum Velocity Comparing SUMO Default, Our Approach (All), Our Approach (Fast), SimRa (All) and SimRa (Fast). All values are in m/s.}%
\label{tab:results_overview_vel}
\begin{tabular}{cccc}
\toprule
& Mean & Std. Deviation & Median\\
\midrule
\midrule
SUMO default & 5.555 & 0.002 & 5.556 \\
\midrule
Our approach (\textit{all}) & 6.589 & 1.234 & 6.58 \\
SimRa (\textit{all}) & 7.072 & 0.904 & 7.10 \\
\midrule
Our approach (\textit{f}) & 8.272 & 0.902 & 8.20 \\
SimRa (\textit{f}) & 7.872 & 0.738 & 7.86 \\
\bottomrule&
\end{tabular}
\end{table}
According to \cref{tab:results_overview_vel} the maximum velocity aspect of our simulation is the closest, when compared to maximum acceleration and maximum deceleration of all and fast (\textit{f}) cyclist groups in SimRa and our approach.
This is probably due to the fact, that we splitted the cyclist based on the average velocity.
Our new parameterization is thus significantly closer to the real-world behavior of cyclists.
As for acceleration and deceleration behaviors, the fact that our new parameterization is not a perfect fit to the real-world data indicates that there are likely to be additional influence factors not captured in our model.

\subsection{Left-turn Behavior at Intersections}%
\label{sec:eval_turn}

\begin{figure*}[t]
    \centering
    \subfloat[Only cyclist group slow ($s$)]{\includegraphics[width=.3\textwidth]{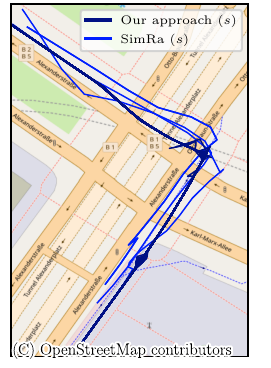}\label{fig:eval_im_traj_slow}}
    \hfill
    \subfloat[Only cyclist group medium ($m$)]{\includegraphics[width=.3\textwidth]{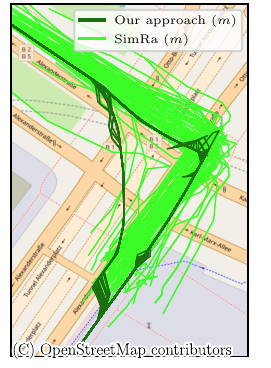}\label{fig:eval_im_traj_medium}}
    \hfill
    \subfloat[Only cyclist group fast ($f$)]{\includegraphics[width=.3\textwidth]{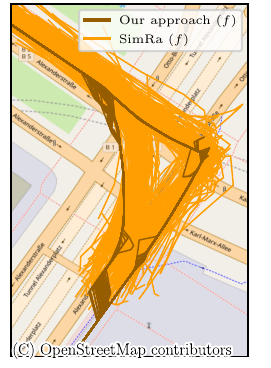}\label{fig:eval_im_traj_fast}}
    \caption{%
        Qualitative comparison between the results of our approach for the intersection model and real-world raw GPS data given by SimRa for the intersection between Alexanderstraße and Karl-Marx-Allee in Berlin.
        SimRa shows two distinct left-turn paths (i.e., a direct and an indirect one), which are also modeled by our models.
        Also the three different groups have different shares between direct and indirect left-turns.
    }%
    \label{fig:eval_im_traj_new}
\end{figure*}

As shown in \cref{fig:eval_im_traj_new}, which shows the intersection between Alexanderstraße and Karl-Marx-Allee in Berlin, the 2D trajectories produced by the new intersection models converge towards the trajectories of the SimRa data set.
While the trajectories produced by SUMO's default bicycle model, as can be seen in \cref{fig:analysis_im_traj_default}, only offer direct "bike-lane-to-bike-lane" turns, the new model is significantly closer to real-world intersection behavior of cyclists.
This is true for all of our cyclist models.
Here, we see again the different left-turn behaviors of different cyclist types.
While slow cyclist only prefer the indirect left-turn, the medium and fast cyclists are more inclined to take the direct left-turn comparatively.

\subsection{Combining Intersection Model and Kinematic Extensions}%
\label{sec:eval_combined}

To achieve a holistic comparison between SUMO's default bicycle model and our new models, we measure the durations of left-turn maneuvers at multiple intersections and compare the empirical distributions of these measurements.
To specifically monitor the impact of our changes, we do not include any ride time before or after the intersection in the measurements.
Note that we also omit the slow rides, since no or too few slow rides went through the intersections presented here.

Based on this, we identified the following four findings:

\begin{figure}[t]
    \centering
    \includegraphics[width=\columnwidth]{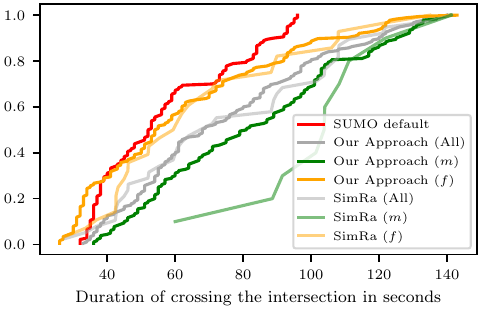}
    \caption{%
        ECDFs of the measured durations for crossing the scenario \textit{Warschauer Straße}.
        It is apparent that our models for all, medium ($m$) and fast ($f$) cyclists outperform SUMO's default as the measured durations converge towards the real-world data.
        Only our fast cyclist model ($f$) is just marginally better than SUMO's default.
    }%
    \label{fig:im_warschauer}
\end{figure}
First, our new models outperform the default at most intersections, as its measured durations converge with real data, see for example \cref{fig:im_warschauer}.
Especially when given the option to use the \textit{indirect} path, cyclists take longer to cross an intersection as they need to stop at an additional traffic light.
This is consistent with real-world data as we find it in the SimRa dataset at multiple intersections.

\begin{figure}[t]
    \centering
    \includegraphics[width=\columnwidth]{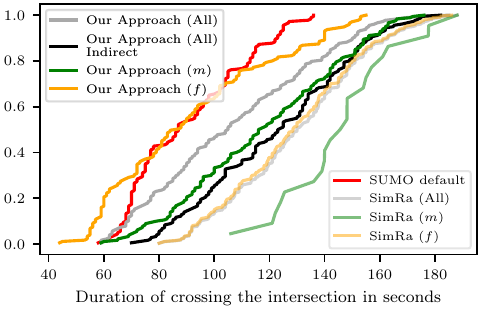}
    \caption{%
        ECDFs of the measured durations for crossing the scenario \textit{Mehringdamm}.
        The results when using our models for all, medium ($m$) and fast ($f$) cyclists are only slightly more realistic than when using the standard SUMO model.
        However, when the \textit{direct} path is blocked for cyclists, the simulation results outperform the default approach.
    }%
    \label{fig:im_mehringdamm}
\end{figure}

Second, in some cases, we were able to improve our results by adjusting the left-turn behavior distribution following the second distribution discussed in \cref{sec:analysis_pathfinding}.
The ``lane only'' results in \cref{fig:im_mehringdamm} were achieved by prohibiting cyclists from using the \textit{direct} path.
Obviously, it takes much longer for cyclists to cross the intersection than SUMO's default simulation model suggests.
When examining SimRa trajectories at this particular intersection, almost all cyclists chose the \textit{indirect} path as the infrastructure guides cyclists to do so.
Hence, manually adjusting the left-turn behavior distribution for such intersections is crucial.

\begin{figure}[t]
    \centering
    \includegraphics[width=\columnwidth]{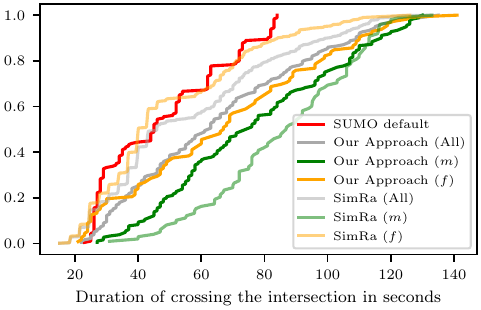}
    \caption{%
        ECDFs of the measured durations for crossing the scenario \textit{Alexanderstraße}.
        Here, our new approach is not more precise than SUMO default behavior.
    }%
    \label{fig:im_alex}
\end{figure}

Third, at a few intersections, the results of our approach do not yet sufficiently reflect real-world bicycle behavior (see \cref{fig:im_alex}).
We discuss possible reasons for this in \cref{sec:disc}.

Fourth, creating three cyclist models based on their average velocity also revealed, that SUMO's default model behaves like fast cyclists.
This becomes clear when looking at \cref{fig:im_warschauer,fig:im_mehringdamm,fig:im_alex}, since the lines representing the cyclists modeled with our fast cyclist model are the closes to the red line representing SUMO default.

\section{Discussion}
\label{sec:disc}
Overall, the results presented in this paper show a significant improvement over the state-of-the-art.
Nevertheless, they still have a number of shortcomings.
In this section, we discuss the inherent limitations of our approach in general (\cref{sec:method}) as well as problems resulting from the SimRa dataset as our ground truth data (\cref{sec:problem_simra}).
We also describe a potential problem regarding e-bikes in \cref{sec:e-bikes}.

\begin{figure}
\includegraphics[width=\columnwidth]{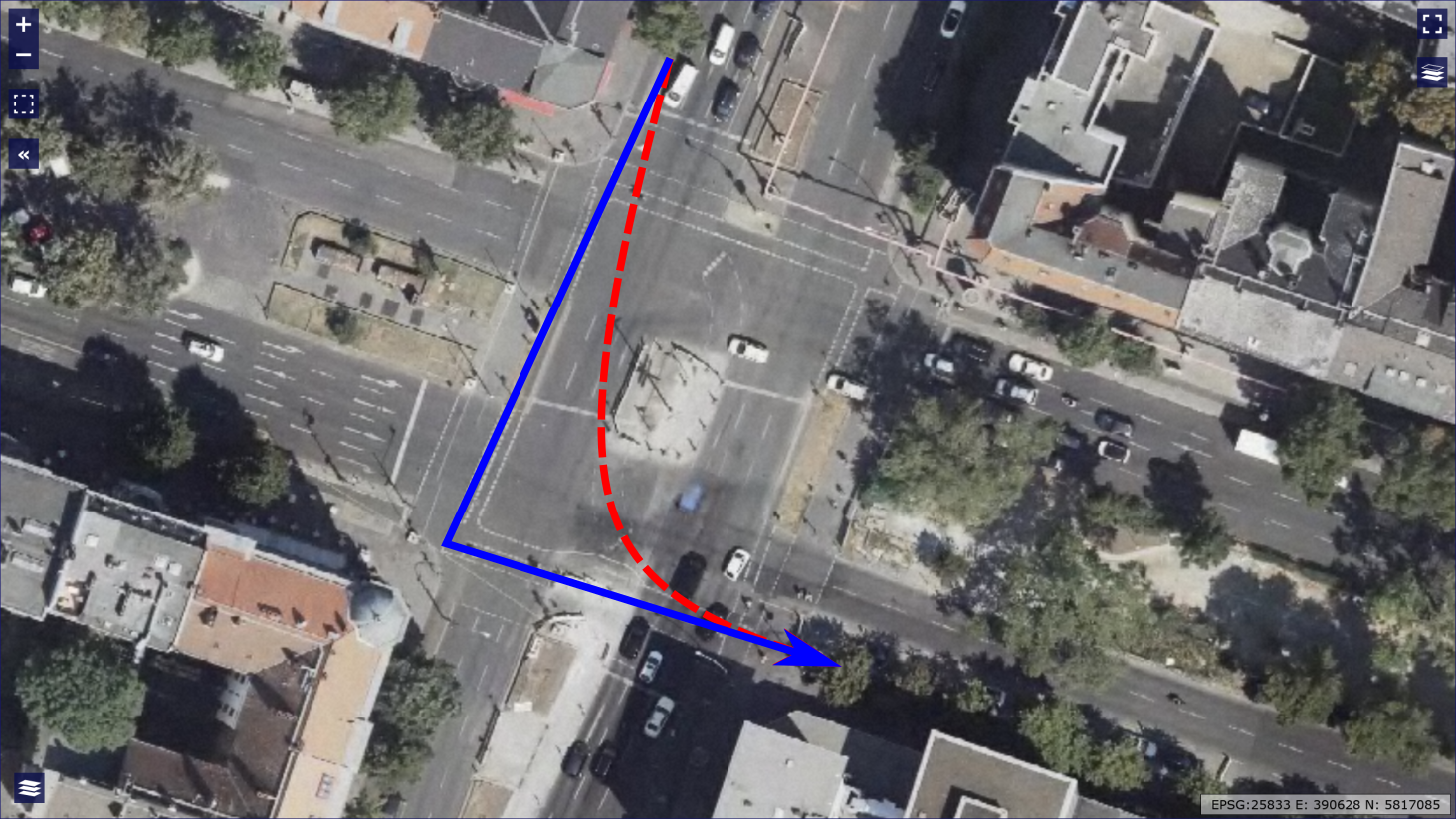}
\caption{Intersection Mehringdamm/Gneisenaustraße: a traffic island obstructs the direct turn path (dashed line) and, thus, makes the indirect path (solid line) more likely to be used.}
\label{fig:mehringdamm_traffic_island}
\end{figure}

\subsection{Methodological Challenges}%
\label{sec:method}

Our initial assumption was that the behavior of cyclists in a single intersection cannot be generalized to all intersections~\cite{kaths2016integration} but that the average behavior across a large number of intersections will be close enough to cyclists' behavior at arbitrary intersections.
This seems to be true only for a (relatively large) subset of intersections -- apparently, the intersection behavior of cyclists is more heterogeneous than expected.
We believe that this is due to the fact that we averaged across all intersections in our dataset whereas there are apparently different classes of intersections that we did not account for.

Primarily, the intersection design is likely to have a strong impact:
Consider the example in \cref{fig:mehringdamm_traffic_island} where a traffic island partially blocks direct left turns and where markings on the ground suggest indirect turns.
As another example, the intersection Bismarckstraße/Leibnizstraße had no direct left turns in the SimRa dataset.
In this intersection, the reason would be that cyclists legally have to use a bike lane.
When using that bike lane, a direct left turn would require cyclists to first pass through a row of parked cars, then to cross four car lanes of a major street before being able to turn left.

Aside from that, other possible influence factors include the amount and velocity of traffic (higher numbers of cars or faster cars can be expected to lead to more indirect turns), gender and age group distributions of cyclists in the respective intersections, as well as weather and light conditions or the grade of the street.
In future work, we plan to explore these possible influence factors, focusing on the intersection design which we deem to have the strongest impact.

Another problem results from inaccuracy in the dataset used:
GPS and motion sensors of smartphones provide only imprecise insights into actual ``micro''-behavior of cyclists.
The problem with GPS sensors on smartphones in our case is that they are not precise enough to detect evasive maneuvers of cyclists, that, e.g., swerve around a pothole, since the GPS modules in smartphones provide location information with an inaccuracy of about 7-13 meters~\cite{merry2019smartphone}.
While human activity recognition based on motion sensors can differentiate very distinct activities such as walking, sitting, or sleeping, this is very challenging in our dataset.
The reasons for this are that (i) we do not have detailed information about the positioning of the phone (which might be next to the knee in case of someone wearing cargo pants or on the bike handlebar), i.e., the data might be extremely noisy or not, (ii) the movements of the cyclist create lots of semi-periodic movements and hence acceleration in all directions, and (iii) the road surface quality can create significant amounts of noise (e.g., a phone mounted on the handlebar while going on a cobblestone road will often experience a maxed-out acceleration sensor for at least one direction).
These all together imply a poor signal-to-noise ratio, i.e., it is very hard to impossible to use the phones' motion sensors to improve GPS quality~\cite{dernbach2012simple,casilari2022an,yang2015deep,karakaya2022cyclesense}.
We tried to partially address these limitations through the combination of preprocessing and using large numbers of rides.
Alternatives would be additional sensors (especially cameras) on bicycles or on intersections~\cite{kaths2016integration}.
These, however, have the inherent limitation that they will either limit the number of bicycles producing data or the number of intersections covered.

\subsection{Dataset Choice as Ground Truth}%
\label{sec:problem_simra}

In this paper, we used the SimRa datasets as input for our analysis as it is, to the best of our knowledge, the first public dataset comprising a large number of rides that actually publishes individual rides in an anonymized but non-aggregated form.
We need to keep in mind, however, that SimRa was designed for a different purpose:
For example, the SimRa app records motion sensors at 50 Hz but only persists every fifth value of a moving average over 30 values.
While this suffices for detecting near miss incidents~\cite{karakaya2020simra}, it further limits the resolution of motion data (and thus any conclusions we can draw from that).
Furthermore, the SimRa data which we used were recorded over a period of 1.5 years.
During such as long period of time, physical changes to the bicycle infrastructure (both temporary and permanent) will occur, thus, adding additional noise to the data.

Finally, SimRa relies on crowdsourcing as a data collection method which often leads to participation inequality.
As a result, individual users will be overrepresented in some intersections and street segments and not represented in others.
Furthermore, based on the data collection method using smartphones, the user group of SimRa is likely to have a slight gender bias towards males and an age group bias towards cyclists between the ages 20 and 50.
These biases will, of course, be reflected in our analysis results and cannot be compensated unless other cycling datasets become available in non-aggregated form.

\subsection{Limitations}%
\label{sec:problem_general}
While the SimRa dataset which we used to analyze the longitudinal kinematic behavior contains rides from almost 100 regions from Germany, Switzerland and Austria, which increases the generalizability, the models presented in this work have their limits:
First, the rides were mostly recorded in highly urban areas.
Second, most of our rides are not recreational rides, but commuting rides from/to school/university/work places~\cite{karakaya2020simra}.
Third, the participating regions in the SimRa projects are all in Germany, Switzerland and Austria.
Fourth, our models do not consider hills as most of our data are from mostly flat areas.
This means that researchers simulating urban, central European, bike commute traffic in mostly flat areas will have a much more realistic simulation when using our models compared to SUMO's default models.
When simulating recreational bike traffic in, e.g., a mountainous region in Japan, our model is unlikely to be a good fit.
For such a scenario, we nevertheless believe that our model might be closer to real behavior than SUMO's default which models cyclists with unrealistically high acceleration and deceleration values coupled with really low maximum velocity.

\subsection{E-Bikes}
\label{sec:e-bikes}
Another potential threat to the realism of our approach are e-bikes, since they support the cyclist, potentially leading to higher acceleration and velocity.
To analyze this, we filtered out the e-bikes and found, that the SimRa dataset only contains about 4,000 e-bike rides, which is less than 5\%.
We redid the analyses of the three cyclist groups and found out, that the change in the results is barely noticeable.
And there are not enough e-bike rides to create a dedicated e-bike model with a high credibility.

\section{Conclusion}
\label{sec:conclusion}
Increasing the modal share of cyclists to provide health benefits, alleviate traffic congestion, and reduce air pollution requires significant planning efforts of city planners and traffic engineers towards an improved cycling infrastructure.
For this, city planners often rely on the open source simulation platform SUMO to study the effects of infrastructure changes before implementing them on the streets.
Likewise, research on V2X-based safety systems for cyclists often relies on SUMO for evaluation.
Unfortunately, SUMO cyclists are either modeled as slow cars or as fast pedestrians, neither of which is overly realistic.

In this paper, we used the recently published SimRa dataset, which to our knowledge is the first public dataset providing detailed insights into a large number of individual cyclists' rides, to improve SUMO's cyclist model.
For this, we split the rides into three categories based on their average velocity, after the stops are filtered out: slow, medium, and fast.
We then derived acceleration, deceleration and velocity behaviors for each of these three cyclist groups and reparameterized the SUMO cyclist models.
As a SUMO extension, we also developed a new intersection model describing left-turn behaviors of cyclists of the three new groups in four-way intersections.
While our work significantly improved the existing cyclist model, it is not as realistic as we wanted it to be.
We, hence, discussed a number of research directions which we plan to explore in the near future.

\bibliographystyle{IEEEtran}
\bibliography{bibliography.bib}

\end{document}